\documentstyle[preprint,aps]{revtex}

\begin{document}
\title{Non-Fermi-liquid behavior in the Kondo lattices induced by peculiarities of
magnetic ordering and spin dynamics}
\author{V.Yu.Irkhin$^*$ and M.I.Katsnelson}
\address{Institute of Metal Physics, 620219 Ekaterinburg, Russia}
\maketitle

\begin{abstract}
A scaling consideration of the Kondo lattices is performed with account of
singularities in the spin excitation spectral function. It is shown that a
non-Fermi-liquid (NFL) behavior between two critical values of the bare $s-f$
coupling constant occurs naturally for complicated magnetic structures with
several magnon branches. This may explain the fact that a NFL behavior takes
place often in the heavy-fermion systems with peculiar spin dynamics.
Another kind of a NFL-like state (with different critical exponents) can
occur for simple antiferromagnets with account of magnon damping, and for
paramagnets, especially with two-dimensional character of spin fluctuations.
The mechanisms proposed lead to some predictions about behavior of specific
heat, resistivity, magnetic susceptibility, and anisotropy parameter, which
can be verified experimentally.
\end{abstract}


\section{Introduction}

Recently, a great experimental material has been obtained for anomalous $f$%
-systems demonstrating so-called non-Fermi-liquid (NFL) behavior \cite
{Maple,Proc}. Manifestations of the NFL behavior are unusual temperature
dependences of magnetic susceptibility ($\chi (T)\sim T^{-\zeta },\zeta <1$%
), electronic specific heat ($C(T)/T$ is proportional to $T^{-\zeta }$ or $%
-\ln T$) and resistivity ($\rho (T)\sim T^\mu ,\mu <2$), etc. Such a
behavior is observed not only in alloys where disorder is present (U$_x$Y$%
_{1-x}$Pd$_3$, UPt$_{3-x}$Pd$_x$, UCu$_{5-x}$Pd$_x$, CeCu$_{6-x}$Au$_x$, U$_x
$Th$_{1-x}$Be$_{13}$), but also in some stoichiometric compounds, e.g., Ce$_7
$Ni$_3$ \cite{CeNi}, CeCu$_2$Si$_2,$CeNi$_2$Ge$_2$ \cite{Steg}. The latter
situation is most interesting from the physical point of view \cite{Alt}.
There are a number of theoretical mechanisms proposed to describe the NFL
state: two-channel Kondo scattering \cite{Tsv,twoch,Col}, ``Griffiths
singularities'' in disordered magnets \cite{Castr,And}, strong spin
fluctuations near a quantum magnetic phase transition \cite{Jul,Ioffe,Col1}
etc. Most of modern treatments of the NFL problem have a
semiphenomenological character. The only microscopic model where formation
of the NFL state is proven - the one-impurity two-channel Kondo model -
seems to be insufficient, since important role of intersite interactions is
now a matter of common experience \cite{Maple}.

In the present paper we start from the standard microscopic model of a
periodical Kondo lattice. Main role in the physics of the Kondo lattices
belongs to the interplay of the on-site Kondo screening and intersite
exchange interactions. This interplay results in the mutual renormalization
of the characteristic energy scales: the Kondo temperature $T_K$ and
spin-fluctuation frequency $\overline{\omega }$. We shall demonstrate that
during the renormalization process ``soft'' boson branches can be formed,
presence of singularities of spin spectral function being of crucial
importance. Scattering of electrons by such soft collective excitations just
leads to the formation of the NFL state (cf. \cite{Col1}).

In Sect. 2 the renormalization group (scaling) equations are presented. In
Sect.3 we consider the antiferromagnetic (AFM) state with account of the
spin-wave damping and the paramagnetic state with simple spin-diffusive
dynamics. It turns out that in these cases a NFL-like behavior (in a
restricted temperature interval) is possible, especially in the case of
quasi-two-dimensional (2D) spin fluctuations. In Sect. 4 we consider the
general problem of singularities of the scaling function. In Sect.5 we show
that the NFL behavior up to lowest temperatures can be naturally obtained
provided that we take into account magnon-like excitations in the case of a
complicated spin dynamics. The excitation picture required is characteristic
for real $f$-systems where several excitation branches exist. In Sect.6 we
discuss various physical properties and possible relation to experimental
data.

\section{The scaling equations}

To describe a Kondo lattice, we use the $s-f$ exchange model
\begin{equation}
H=\sum_{{\bf k}\sigma }t_{{\bf k}}c_{{\bf k}\sigma }^{\dagger }c_{{\bf k}%
\sigma }-I\sum_{i\alpha \beta }{\bf S}_i\mbox {\boldmath $\sigma $}_{\alpha
\beta }c_{i\alpha }^{\dagger }c_{i\beta }+\sum_{{\bf q}}J_{{\bf q}}{\bf S}_{%
{\bf -q}}{\bf S}_{{\bf q}}+H_a
\end{equation}
where $t_{{\bf k}}$ is the band energy, ${\bf S}_i$ and ${\bf S}_{{\bf q}}$
are spin-density operators and their Fourier transforms, $I$ is the $s-f$
exchange parameter, $J_{{\bf q}}$ are the intersite exchange parameters, $%
\sigma $ are the Pauli matrices, $H_a$ is the anisotropy Hamiltonian which
results in occurrence of the gap $\omega _0$ in the spin-wave spectrum. In
Refs.\cite{kondo,kondo1}, the interplay of the Kondo effect and intersite
interactions was investigated by the renormalization group approach. The
latter starts from the second-order perturbation theory with the use of the
equation-of-motion method (within the diagram technique in the pseudofermion
representation for the spin operators, such an approximation corresponds to
the one-loop scaling). The relevant variables are the effective
(renormalized) parameter of $s-f$ coupling $g_{ef}(C)=-2\rho I_{ef}(C)$ ($%
C\rightarrow -0$ is a flow cutoff parameter, $\rho $ is the bare density of
electron states at the Fermi level), characteristic ``exchange''
spin-fluctuation energy $\overline{\omega }_{ex}(C)$, gap in the spin-wave
spectrum $\omega _0(C)$, and magnetic moment $\overline{S}_{ef}(C)$. In the
magnetically ordered phase, $\overline{\omega }_{ex}$ is the magnon
frequency $\omega _{{\bf q}},$ which is averaged over the wavevectors ${\bf %
q=}2{\bf k}${\bf \ }where ${\bf k}$ runs over the Fermi surface; in the case
of dissipative spin dynamics (paramagnetic phase) $\overline{\omega }_{ex}$
is determined by the second moment of the spin spectral density.

Here we write down the set of scaling equations with account of the magnon
damping $\overline{\gamma }(C)$:
\begin{eqnarray}
\partial g_{ef}(C)/\partial C &=&\Lambda ,\partial \ln \overline{S}%
_{ef}(C)/\partial C=-\Lambda /2  \label{gl} \\
\partial \ln \overline{\omega }_{ex}(C)/\partial C &=&-a\Lambda /2,\partial
\ln \omega _0(C)/\partial C=-b\Lambda /2  \label{sl} \\
\partial \ln \overline{\gamma }(C)/\partial C &=&-c\Lambda /2  \label{gaml}
\end{eqnarray}
with
\[
\Lambda =\Lambda (C,\overline{\omega }_{ex}(C),\omega _0(C))=\frac{%
g_{ef}^2(C)}{|C|}\eta \left( \frac{\overline{\omega }_{ex}(C)}{|C|},\frac{%
\omega _0(C)}{|C|},\frac{\overline{\gamma }(C)}{|C|}\right) ,
\]
$a=1-\alpha $ for the paramagnetic (PM) phase, $a=1-\alpha ^{\prime },b=1$
for the antiferromagnetic (AFM) phase, $a=2(1-\alpha ^{\prime \prime }),b=2$
for the ferromagnetic (FM) phase; $\alpha ,\alpha ^{\prime },\alpha ^{\prime
\prime }$ are some averages over the Fermi surface (see Ref.\cite{kondo}).
For the staggered AFM ordering in the 2D and 3D cubic lattices where
\[
J_{{\bf q}}=2J_1\sum_{i=1}^d\cos q_i+4J_2\sum_{i>j}^d\cos q_i\cos q_j
\]
with $J_1$ and $J_2$ being the exchange integrals between nearest and
next-nearest neighbors ($|J_1|\gg |J_2|$), one can derive by using expansion
in small $q$%
\begin{equation}
\alpha ^{\prime }\simeq 2(d-1)\frac{J_2}{J_1}\left| \left\langle \exp (i{\bf %
kR}_2)\right\rangle _{t_{{\bf k}}=0}\right| ^2  \label{alph}
\end{equation}
where ${\bf R}_2$ runs over the next-nearest neighbors, the electron
spectrum $t_{{\bf k}}$ is referred to the Fermi energy $E_F=0$. Similar
equations can be obtained for the Coqblin-Schrieffer model \cite{kondo} and
in the case of anisotropic $s-f$ coupling \cite{kondo1}.

Using (\ref{gl})-(\ref{gaml}) we obtain the explicit expressions
\begin{eqnarray}
\overline{\omega }_{ex}(C) &=&\overline{\omega }_{ex}\exp (-a[g_{ef}(C)-g]/2)
\label{w+g} \\
\omega _0(C) &=&\omega _0\exp (-b[g_{ef}(C)-g]/2)  \nonumber \\
S_{ef}(C) &=&S\exp (-[g_{ef}(C)-g]/2)  \nonumber \\
\overline{\gamma }(C) &=&kg_{ef}^2(C)\overline{\omega }\exp
(-c[g_{ef}(C)-g]/2)  \nonumber
\end{eqnarray}
In the last equation of (\ref{w+g}), we have taken into account that the
magnon damping is proportional to $g^2,$ $\overline{\gamma }=kg^2\overline{%
\omega }$ (the factor $k$ is determined by the bandstructure and magnetic
ordering), and the scaling equations should contain only renormalized
quantities.

\section{The scaling behavior in the presence of damping}

As demonstrated in Ref.\cite{kondo}, the singularities of the scaling
function $\eta $ can result in occurrence of a NFL behavior in a restricted
region due to fixing of the argument of the function $\eta $ at the
singularity during the scaling process, so that $\overline{\omega }(C)\simeq
|C|,$%
\begin{equation}
g_{ef}(\xi )-g\simeq 2(\xi -\lambda )/a,\lambda \equiv \ln (D/\overline{%
\omega }),  \label{linnn}
\end{equation}
(the scale $D$ is defined by $g_{ef}(-D)=g,\xi \equiv \ln |D/C|$). This
region becomes not too narrow only provided that the bare coupling constant $%
g=-2I\rho $ is very close to the critical value $g_c$ for the magnetic
instability ($|g-g_c|/g_c\sim 10^{-4}\div 10^{-6}$). Here we consider the
scaling process with account of a not too small magnon damping (only very
small damping was introduced in Ref.\cite{kondo} to provide existence of the
magnetic-non-magnetic ground-state transition at $g=g_c$).

The scaling function $\eta $ in the FM and AFM phases for simple magnetic
structures reads (for simplicity, magnetic anisotropy is neglected in this
Section)

\begin{equation}
\eta \left( \overline{\omega }_{ex}/|C|,\overline{\gamma }/|C|\right) =\text{%
Re}\left\langle \left( 1-(\omega _{{\bf k-k}^{\prime }}^{}+i\gamma _{{\bf k-k%
}^{\prime }}^{})^2/C^2\right) ^{-1}\right\rangle _{t_k=t_{k^{\prime }}=0}
\label{etafm}
\end{equation}
For an isotropic three-dimensional (3D) ferromagnet integration in (\ref
{etafm}) for $\gamma =$ const and quadratic spin-wave spectrum yields
\begin{equation}
\eta (x,z)=\frac{1+z^2}{4x}\ln \frac{(1+x)^2+z^2}{(1-x)^2+z^2}+\frac{1+x}z%
\arctan \frac z{1+x}-\frac{1-x}z\arctan \frac z{1-x}  \label{intfm}
\end{equation}
where $x=\overline{\omega }_{ex}/|C|,z=$ $\overline{\gamma }/|C|.$ Note that
last two terms in (\ref{intfm}) play a role similar to that of the
``incoherent'' contribution to the function $\eta ,$ which was treated in
Ref.\cite{kondo}.

For a 3D antiferromagnet integration in (\ref{etafm}) for the linear
spin-wave spectrum gives
\begin{equation}
\eta (x,z)=\frac 12\text{Re }[(1+iz)\ln (1+x+iz)+(1-iz)\ln (1+x-iz)]
\end{equation}
where we take into account the intersubband damping only, 
\begin{eqnarray}
\gamma _{{\bf q}} &=&\pi 2I^2\overline{S}(J_{{\bf q}}-J_{{\bf Q}})\rho
^2\lambda _{{\bf q+Q}},  \label{dintr} \\
\lambda _{{\bf q}} &=&\rho ^{-2}\sum_{{\bf k}}\delta (t_{{\bf k}})\delta (t_{%
{\bf k+q}}).
\end{eqnarray}
The damping (\ref{dintr}) can be put nearly constant at not too large $q$
(the threshold value determined by the AFM gap can be neglected due to
formal smallness in $I$). In the 2D case we obtain in the same approximation 
\begin{eqnarray}
\eta (x,z) &=&\frac{\nu ^3(x,z)}{\nu ^4(x,z)+z^2} \\
\nu ^2(x,z) &=&\frac 12[1-x^2+\sqrt{(1-x^2)^2+4z^2}]  \nonumber
\end{eqnarray}
which modifies somewhat the result of Ref.\cite{kondo}.

The function $\lambda _{{\bf q}}$ determines, in particular, the factor $k$
in (\ref{w+g}). For a parabolic electron spectrum we obtain 
\begin{equation}
\lambda _{{\bf q}}=\frac{\theta (1-x)}{zx}\times \left\{ 
\begin{tabular}{ll}
$1/6,$ & $d=3$ \\ 
$(4\pi )^{-1}(1-x^2)^{-1/2},$ & $d=2$%
\end{tabular}
\right.  \label{lam}
\end{equation}
where $x=q/2k_F,$ $\theta (x)$ is the step function, $z$ is the electron
concentration (with both spin projections).

Main Kondo renormalization of the magnon damping comes from its
proportionality to the factor of $\overline{S}.$ Spin fluctuations can give
correction to this factor, as well as for the magnon frequency \cite{kondo}.
For simplicity, we restrict ourselves in numerical calculations to the AFM
case with $\delta \gamma _{{\bf q}}/\gamma _{{\bf q}}=\delta \omega _{{\bf q}%
}/\omega _{{\bf q}}=\delta \overline{S}/S,$ so that $c=1$ in (\ref{gaml}), (%
\ref{w+g}). The corresponding scaling trajectories are shown in Fig.1. One
can see that, unlike Ref.\cite{kondo}, the ``linear'' behavior, although
being somewhat smeared, is pronounced in a considerable region of $\xi $ for
not too small $|g-g_c|$, especially in the 2D case. In the 3D case the
linear region (\ref{linnn}) is followed by a quasi-linear behavior with 
\begin{equation}
g_{ef}(\xi )\simeq A(\xi -\lambda ),\overline{\omega }(C)\propto |C|^{aA/2},%
\overline{S}_{ef}(C)\propto |C|^{A/2},  \label{lina}
\end{equation}
where $A<2/a.$ To investigate the latter behavior in more details, it is
instructive to consider also the case of a paramagnet with pure dissipative
dynamics. In the case of spin-diffusion behavior we have (cf. Ref.\cite
{kondo}) 
\begin{equation}
\eta ^{PM}(\frac{\overline{\omega }}C)=\sum_{{\bf q}}\lambda _{{\bf q}%
}\left[ 1+({\cal D}q^2/C)^2\right] ^{-1},\overline{\omega }=4{\cal D}k_F^2
\label{etpm}
\end{equation}
where ${\cal D}$ is the spin diffusion constant. As demonstrate numerical
calculations (see Fig.2), for $g\leq g_c$ the one-impurity behavior $%
1/g_{ef}(\xi )=1/g-\xi $ is changed at $\xi \simeq \lambda $ by the behavior
(\ref{lina}) with 
\begin{equation}
A\simeq [g_{ef}(\lambda )]^2\Psi (0),g_{ef}(\lambda )\simeq g/(1-\lambda
g),\Psi (0)=\eta (1)\sim 0.5  \nonumber
\end{equation}
In the 3D case where $\eta ^{PM}(x)=\arctan x/x$ the quasi-linear NFL-like
behavior $g_{ef}(\xi )$ takes place in a rather narrow region. However, in
the 2D case we obtain 
\begin{equation}
\eta ^{PM}(x)=\left[ \frac{1+(1+x^2)^{1/2}}{2(1+x^2)}\right] ^{1/2}
\end{equation}
and the NFL-like region becomes more wide due to a more slow decrease of $%
\eta ^{PM}(x)$ at $x\rightarrow \infty .$ Note that such regions are not
observed for $g>g_c.$

\section{Singularities of the scaling function}

To get further insight into the NFL-behavior problem, we perform an analysis
of singularities of the scaling function $\eta ,$ which is more general than
in Ref.\cite{kondo}. In the absence of damping we can write down

\begin{equation}
\eta \left( \overline{\omega }_{ex}/|C|,\omega _0/|C|\right) =\sum_{{\bf q}%
}\lambda _{{\bf q}}\left( 1-\omega _{{\bf q}}^2/C^2\right) ^{-1}
\label{etafm1}
\end{equation}
The singularities of $\eta $ correspond to the Van Hove singularities in the
magnon spectrum and to the boundary points $q=0$ and $q=2k_F.$

For $q\rightarrow 2k_F$ the magnon spectrum has the Kohn anomaly 
\begin{equation}
\delta \omega _{{\bf q}}\propto \left\{ 
\begin{tabular}{ll}
$(q-2\dot k_F)\ln |q-2k_F|,$ & $d=3$ \\ 
$\sqrt{q^2-4k_F^2}\theta (q-2k_F),$ & $d=2$%
\end{tabular}
\right.
\end{equation}
Taking into account the dependences (\ref{lam}), which hold qualitatively in
a general situation, we obtain for $v=C^2-\overline{\omega }^2\rightarrow 0$%
\begin{equation}
\eta (v)\propto \left\{ 
\begin{tabular}{ll}
$\ln \ln |v|,$ & $d=3$ \\ 
$\theta (v)v^{-1/2},$ & $d=2$%
\end{tabular}
\right.
\end{equation}
Note that the singularity of the form $\ln v$ obtained for the $3D$ case in
Ref.\cite{kondo} (see also previous Section), is a consequence of a
simplified ``Debye'' model of the magnon spectrum. In fact, such a
dependence corresponds qualitatively to an intermediate asymptotics at
approaching the singularity. In any case, a small damping of spin
excitations should be introduced to cut the singularity. In the calculations
below we put the damping parameter $\delta =1/100$ (see Ref.\cite{kondo}).
In fact, pronounced extrema of $\eta ,$ but not singularities themselves
turn out to be important for the scaling behavior discussed below.

For $q\rightarrow 0$ we have $\lambda _{{\bf q}}\propto q^{-1}.$ Near the
points of minimum (maximum) in the magnon spectrum we have $\omega _{{\bf q}%
}^2-\omega _m^2\propto \pm q$ $^2,$ and for $v=C^2-\omega _m^2\rightarrow 0$
we obtain 
\begin{equation}
\eta (v)\propto \left\{ 
\begin{tabular}{ll}
$\pm \ln |v|,$ & $d=3$ \\ 
$\mp \theta (\mp v)|v|^{-1/2},$ & $d=2$%
\end{tabular}
\right.
\end{equation}
so that $\eta \rightarrow -\infty $ near the band bottom and $\eta
\rightarrow +\infty $ near the band top. The Van Hove singularities in the
magnon band at $\omega =\omega _c$ for $q\neq 0$ yield weaker singularities
in $\eta (v)$ ($|v|^{1/2}$ for $d=3$ and a finite jump for $d=2$) and will
not be treated below.

\section{The scaling picture and NFL behavior in many-sublattice magnets}

Using the results of the previous Section we can propose a rather realistic
and universal mechanism of the NFL behavior. Suppose that the spin
excitation spectrum contains several branches which make additive
contributions to the function $\eta .$

As a simple example we can consider a two-sublattice ferrimagnet with the
localized-system Hamiltonian 
\begin{equation}
H_f=\sum_{{\bf q}}(J_{{\bf q}}{\bf S}_{{\bf -q}}{\bf S}_{{\bf q}}+J_{{\bf q}%
}^{\prime }{\bf S}_{{\bf -q}}^{\prime }{\bf S}_{{\bf q}}^{\prime }+%
\widetilde{J}_{{\bf q}}{\bf S}_{{\bf -q}}{\bf S}_{{\bf q}}^{\prime }),
\end{equation}
the $s-f$ exchange interaction being taken into account only at one
sublattice (spins without primes). Similar to (\ref{etafm1}) we obtain 
\begin{equation}
\eta =\sum_{{\bf q,}i=1,2}\lambda _{{\bf q}}\frac{\omega _{{\bf q}i}}{B_{%
{\bf q}}}\left( 1-\frac{\omega _{{\bf q}i}^2}{C^2}\right) ^{-1}
\end{equation}
where 
\begin{eqnarray}
B_{{\bf q}} &=&\{[S(J_{{\bf q}}-J_0)+S^{\prime }(J_{{\bf q}}^{\prime
}-J_0^{\prime })-(S+S^{\prime })\widetilde{J}_0]^2-4SS^{\prime }\widetilde{J}%
_{{\bf q}}^2\}^{1/2}, \\
\omega _{{\bf q}1,2} &=&B_{{\bf q}}\mp |S(J_{{\bf q}}-J_0)-S^{\prime }(J_{%
{\bf q}}^{\prime }-J_0^{\prime })+(S-S^{\prime })\widetilde{J}_0|  \nonumber
\end{eqnarray}
are the acoustical and optical modes.

The dependence 
\begin{equation}
\Psi (\xi )=\sum_iz_i\eta _i[(\overline{\omega }_{ex,i}/D)e^\xi ,(\omega
_{0,i}/D)e^\xi ]  \label{psiop}
\end{equation}
is shown in Fig.3 for the case of two excitation modes in a 3D
antiferromagnet (the expressions for the function $\eta _i$ in the case of
one mode with inclusion of anisotropy are given in Ref.\cite{kondo1}). The
only property of the function $\Psi $, which will be important below, is the
occurrence of the {\it second }zero with decreasing $|C|$ (or increasing $%
\xi $). This property follows immediately from existence of the ``positive''
singularity in $\Psi $ near the maximum frequency of the lower branch and of
the ``negative'' singularity near the minimum frequency of the upper branch
(e.g., for $\omega _1(q=2k_F)<\omega _2(q=0)$). One can expect that this is
a general property of many-sublattice magnets. The singularities can be also
connected with the crystal-field excitations \cite{FLow}.

As demonstrate both numerical calculations and analytical treatment, in some
interval of the bare coupling parameter, $g_{c1}<g<g_{c2},$ the argument of
the function $\Psi $ becomes fixed at the second zero, $C=C_0,$ during the
scaling process which is described by Eq.(\ref{gl}). This can be illustrated
for the simple case where $a=b$ for all the modes (e.g., for an
antiferromagnet in the nearest-neighbor approximation we have $a=b=1$). On
substituting (\ref{w+g}) into (\ref{gl}) we obtain 
\begin{equation}
\partial (1/g_{ef})/\partial \xi =-\Psi (\xi -a[g_{ef}-g]/2)  \label{gl1}
\end{equation}
Then we derive 
\begin{equation}
g_{ef}(\xi \rightarrow \infty )\simeq (2/a)(\xi -\xi _0)-1/[\Psi ^{\prime
}(\xi _0)\xi ^2]  \label{lin}
\end{equation}
where $\xi _0\sim \ln (D/\overline{\omega })$ is the second zero of the
function $\Psi (\xi ).$ Note that the first zero does not work since the
unrestricted increase of $g_{ef}$ corresponds to a decrease of the argument
of the function $\Psi $ in (\ref{gl1}), so that $\Psi \rightarrow +0.$
According to (\ref{w+g}) we obtain 
\begin{equation}
\overline{\omega }_{ex}(C),\omega _0(C)\propto |C|,\overline{S}%
_{ef}(C)\propto |C|^{1/a}  \label{lin1}
\end{equation}
Within the approach used, the behavior (\ref{lin}),(\ref{lin1}) takes place
up to $C=0$ $(\xi =\infty ).$ Although the one-loop scaling equations
themselves may become invalid with increasing $g_{ef},$ the tendency to the
formation of the ``soft'' magnon mode seems to be physically correct. The
scaling picture for three possible cases is shown in Fig.4. One can see that
the interval $[g_{c1},g_{c2}]$ where the NFL behavior occurs is not too
small, unlike Ref.\cite{kondo}.

In a more general case, where $a\neq b$ and the exponents in (\ref{sl})
differ for different frequencies, the linear $C$-dependence takes place only
for the total characteristic frequency $\overline{\omega }$ (e.g., for an
anisotropic antiferromagnet with one mode we have $\overline{\omega }%
^2=\omega _0^2+\overline{\omega }_{ex}^2$), and the behavior $g_{ef}(\xi )$
and $\overline{S}_{ef}(\xi )$ is more complicated. As follows from (\ref
{alph}), for the AFM state with small next-nearest exchange interactions the
case of small $|a-1|$ and $b=$ $1$ is realized.

\section{Discussion of physical properties and conclusions}

Consider the temperature dependence of the magnetic susceptibility. In the
spin-wave region we have for an AFM structure with the wavevector ${\bf Q}$%
\begin{equation}
\chi =\lim_{q\rightarrow 0}\langle \langle S_{{\bf q}}^x|S_{-{\bf q}%
}^x\rangle \rangle _{\omega =0}=(J_0-J_{{\bf Q}})^{-1}\propto \overline{S}/%
\overline{\omega }
\end{equation}
One can assume that the spin-wave description of the electron-magnon
interaction is adequate not only in the AFM phase, but also for systems with
a strong short-range AFM order (e.g., for 2D and frustrated 3D systems at
finite temperatures). Anomalous $f$-systems demonstrate indeed pronounced
quasi-2D spin fluctuations, see, e.g., Refs.\cite{Col1,Stock}. Using the
scaling arguments we can replace $\overline{\omega }\rightarrow \overline{%
\omega }(C),$ $\overline{S}\rightarrow \overline{S}_{ef}(C)$ with $|C|\sim
T, $ which yields 
\begin{equation}
\chi (T)\propto T^{-\zeta },\zeta =(a-1)/a
\end{equation}
According to (\ref{alph}), the non-universal exponent $\lambda $ is
determined by details of magnetic structure and can be both positive and
negative. For a qualitative discussion, we can still use Fig.4 and treat the
difference $a-1$ as a perturbation. The increase of $\chi (T\rightarrow 0)$
(which is usually called NFL behavior) takes place for $a>1$ and, as follows
from (\ref{w+g}), corresponds to an increase of magnetic anisotropy
parameter with lowering $T$ (see Ref.\cite{kondo1}). Such a correlation may
be verified experimentally.

The temperature dependence of electronic specific heat can be estimated from
the second-order perturbation theory, $C_{el}(T)/T\propto 1/Z(T)$ where $%
Z(T) $ is the residue of the electron Green's function at the distance $T$
from the Fermi level (cf. Ref.\cite{IKFTT}). Then we have 
\begin{equation}
C_{el}(T)/T\propto g_{ef}^2(T)\overline{S}_{ef}(T)/\overline{\omega }%
_{ex}(T)\propto \chi (T)\ln ^2T
\end{equation}
The dependence $C_{el}(T)/T\propto \chi (T)$ has been recently obtained
experimentally for a wide class of NFL systems \cite{And}. One can expect
that the accuracy of the experimental data is insufficient to pick out the
factor of $\ln ^2T.$ Generally, the temperature behavior of magnetic
characteristics ($\overline{S}$ and $\overline{\omega }$), which depend
exponentially on the coupling constant, is decisive for our NFL mechanisms.
The transport relaxation rate determining the temperature dependence of the
resistivity owing to scattering by spin fluctuations in AFM phase is given
by \cite{afm} 
\begin{equation}
\frac 1\tau =\frac{2\pi }{v_F^2}I^2\overline{S}^2(J_0-J_{{\bf Q}})\rho
\langle ({\bf v_{k+Q}-v}_{{\bf k}})^2\rangle _{t_{{\bf k}}=0}\sum_{{\bf %
q\simeq Q}}\lambda _{{\bf q}}\left( -\frac{\partial N_{{\bf q}}}{\partial
\omega _{{\bf q}}}\right)
\end{equation}
Then we obtain 
\begin{equation}
\ \frac 1\tau \propto g_{ef}^2(T)\overline{S}_{ef}(T)/\overline{\omega }%
_{ex}(T)T^2\propto T^2C_{el}(T)/T\propto T^{2-\zeta }  \label{taul}
\end{equation}
Considering electron-electron scattering as the main scattering mechanism
one can expect another temperature dependence, namely

\begin{equation}
\frac 1\tau \propto \left[ \rho /Z(T)\right] ^2T^2\propto T^2\left[
C_{el}(T)/T\right] ^2\propto T^{2-2\zeta }  \label{tau2}
\end{equation}
It would be interesting to compare experimental dependences on one hand, of
the resistivity and, on the other hand, of the susceptibility and specific
heat, to choose between dependences (\ref{taul}) and (\ref{tau2}). Validity
of the relation (\ref{taul}) would be a verification of the scenario for NFL
state formation proposed here.

The temperature dependences of specific heat, magnetic susceptibility and
resistivity in the case of a NFL-like behavior considered in Sect.3 differ
from those discussed above by the value of $\zeta =(a-1)A/2$ (for the second
``linear'' region).

To conclude, phenomenon of the NFL behavior seems to have a complicated
nature, so that it is hardly possible to propose an unified picture for all
the cases. The mechanisms considered above are not based on disorder
effects, but describe naturally the NFL state in ideal crystals. At the same
time, the damping makes the quasi-NFL behavior considered in Sect.3 more
pronounced and in a sense plays the role of disorder.

The damping is not important for the NFL behavior mechanism considered in
Sect.5. This NFL picture is ``true'' (i.e. holds up to lowest temperatures)
within the lowest-order scaling approach; treatment of higher-order
corrections to the scaling equations would provide additional information.
Unlike previous phenomenological works, existence of peculiar long-range
critical fluctuations near the quantum phase transition is not needed for
this mechanism, but local reconstruction of electronic states owing to the
Kondo effect is essential, the concrete form of spin spectral function being
of crucial importance. More detailed investigations of the NFL behavior for
complicated spectral functions, in particular with account of incoherent
contributions, would be also of interest.

The research described was supported in part by Grant No.99-02-16279 from
the Russian Basic Research Foundation.

\newpage\ 

{\sc Figure captions}

Fig.1. The scaling trajectories $g_{ef}(\xi )$ in isotropic 2D
antiferromagnets (solid lines, $g=0.154<g_c,g=0.155>g_c$) and 3D
antiferromagnets (dashed lines, $g=0.139<g_c,g=0.140>g_c$) with $%
k=0.5,a=c=1,\lambda =\ln (D/\overline{\omega })=5.$

Fig.2. The scaling trajectories $g_{ef}(\xi )$ in 2D (solid lines) and 3D
(dashed lines) paramagnets with $a=1/2,\lambda =5.$ The bare coupling
parameters are $g=0.135$ and $g=0.145,$ higher curves corresponding to
larger $g.$ The values of $g_c$ in 2D and 3D cases are 0.148 and 0.152.

Fig.3. The scaling functions $\Psi (\xi )$ defined by (\ref{psiop}) in the
case of a 3D antiferromagnet with two excitation modes The parameters are $%
z_1=0.4,z_2=0.6,$ $\ln (D/\overline{\omega }_2)=4,\overline{\omega }_2/%
\overline{\omega }_1=3,$ $\omega _{0,1}/\overline{\omega }_{ex,1}=0.2,\omega
_{0,2}/\overline{\omega }_{ex,2}=0.6$ ($\overline{\omega }_i^2=\overline{%
\omega }_{ex,i}^2+\omega _{0,i}^2$).

Fig.4. The scaling trajectories $g_{ef}(\xi )$ in a 3D antiferromagnet with
the parameters of Fig.3 and $a_i=b_i=1.$ The bare coupling parameters are $%
g=0.158\simeq g_{c2}$ (upper solid line with the asymptotics), $%
g=0.154\simeq g_{c1}$ (lower solid line), and $g_{c1}<g=0.156<g_{c2}$
(dashed line).

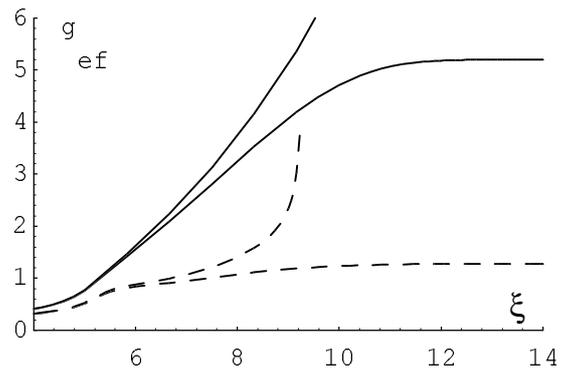
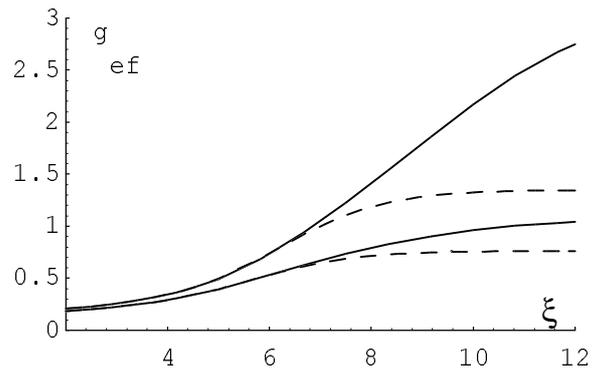
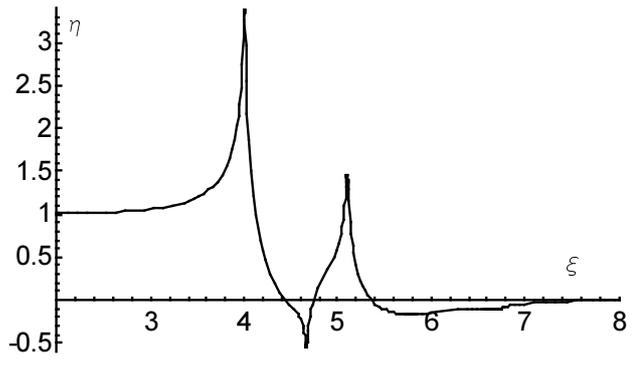
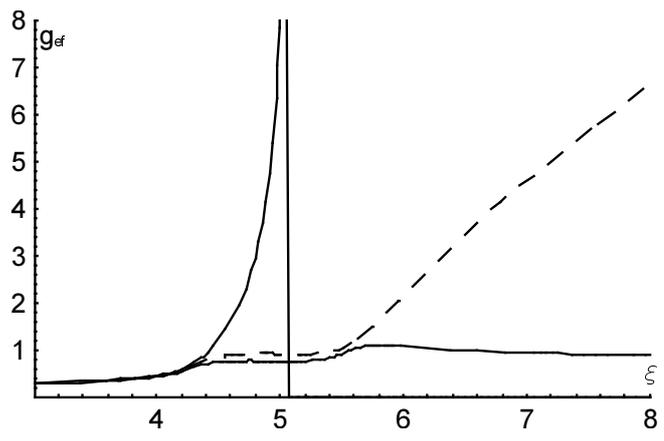
\end{document}